\documentclass[fleqn,usenatbib,useAMS]{mnras}

\usepackage{graphicx}	
\usepackage{amsmath}	
\usepackage{amssymb}	
\usepackage{multicol}        
\usepackage{bm}		
\usepackage{pdflscape}	

\usepackage[T1]{fontenc}
\usepackage{ae,aecompl}

\usepackage{newtxtext,newtxmath}

\title[Non-uniform coronal loops] {\bf Static and dynamic solar coronal loops with cross-sectional area variations}

\author[P.J. Cargill et al]{P. J. Cargill$^{1,}$$^{2}$ \thanks{Contact e-mail: \href{mailto:pcargill@st-andrews.ac.uk}{pcargill@st-andrews.ac.uk}} S. J. Bradshaw$^{3}$ J. A. Klimchuk$^{4}$ W. T. Barnes$^{5}$%
\\
$^{1}$School of Mathematics and Statistics, University of St Andrews, St Andrews, Fife, KY16 9SS, United Kingdom\\
$^{2}$Space and Atmospheric Physics, The Blackett Laboratory, Imperial College, London SW7 2BW, United Kingdom\\
$^{3}$Department of Physics and Astronomy, Rice University, Houston, TX, 77005, USA.\\
$^{4}$Heliophysics Science Division, Goddard Space Flight Center, Greenbelt, MD 20771, USA.\\
$^{5}$National Research Council Postdoctoral Research Associate residing at the Naval Research Laboratory, Washington, D.C. 20375, USA}

\date{Accepted Oct 27 2020}

\pubyear{2020}

\begin{document}
\label{firstpage}
\pagerange{\pageref{firstpage}--\pageref{lastpage}}
\maketitle

\begin{abstract}
{The Enthalpy Based Thermal Evolution of Loops (EBTEL) approximate model for static and dynamic coronal loops is developed to include the effect of a loop cross-sectional area which increases from the base of the transition region (TR) to the corona. The TR is defined as the part of a loop between the top of the chromosphere and the location where thermal conduction changes from an energy loss to an energy gain. There are significant differences from constant area loops due to the manner in which the reduced volume of the TR responds to conductive and enthalpy fluxes from the corona. For static loops with modest area variation the standard picture of loop energy balance is retained, with the corona and TR being primarily a balance between heating and conductive losses in the corona, and downward conduction and radiation to space in the TR. As the area at the loop apex increases, the TR becomes thicker and the density in TR and corona larger. For large apex areas, the coronal energy balance changes to one primarily between heating and radiation, with conduction playing an increasingly unimportant role, and the TR thickness becoming a significant fraction of the loop length. Approximate scaling laws are derived that give agreement with full numerical solutions for the density, but not the temperature.  For non-uniform areas, dynamic loops have a higher peak temperature and are denser in the radiative cooling phase by of order 50\% than the constant area case for the examples considered. They also show a final rapid cooling and draining once the temperature approaches 1 MK. Although the magnitude of the emission measure will be enhanced in the radiative phase, there is little change in the important observational diagnostic of its  temperature dependence.}
\end{abstract}

\begin{keywords}
Sun: corona - Sun: magnetic fields
\end{keywords}



\newpage

\section{Introduction}
The magnetically closed solar corona has been the subject of modelling efforts for almost five decades. The structures observed there, both the easily distinguished loops and the more diffuse background, have a wide range of temperature and brightness, depending on whether they are in the quiet sun or active regions. [For simplicity we refer to all such magnetically closed structures as loops.] In active regions, spatially averaged fairly steady emission with temperatures of up to 3 MK is detected \citep[e.g.][]{warren_2012}. Such structures are assumed to be heated by an as-yet-undetermined process, but almost certainly related to the coronal magnetic field \citep[e.g.][]{reale_2014,klim_rev_2015}. Whether this heating is highly impulsive, or close to being steady, is as yet unknown, but there is in reality almost certainly a continuum of the quantity of energy released in such events \citep[e.g.][]{dem_browning_2015}. Note that the averaged steady emission from active regions is likely to be the integrated signature of many impulsive heating events \citep{cargill2015}.

One common approach to modeling steady and impulsive coronal heating involves solving numerically the one-dimensional hydrodynamic equations along a magnetic field line in response to an imposed heating function \citep[e.g.][]{reale_2014}. The output of such models are the density, temperature and velocity as a function of position and time. In fact, this is very challenging, especially for dynamic models, since the heat flux from the heated corona to the transition region (TR) and upper chromosphere leads to very steep temperature gradients in these lower regions: the temperature scale height, defined as $L_T = T/|dT/ds|$, can be as small as 100 m while the loop can have a length of 100 Mm \citep[e.g.][]{sb_pc2013}. In turn this requires a very fine numerical grid which imposes a severe limit on the timestep in order to ensure stability of the heat conduction solver. A coarse grid leads to major errors in the coronal density arising from the heating \citep{sb_pc2013}, although approximate ways of treating the TR \citep{lionello_2009,mikic_2013,johnston_2017a,johnston_2020} can mitigate this.

An alternative approach is to use approximate methods for solving the coronal hydrodynamic equations: early work was reviewed by \citet{ebtel_2012b} and over the last 15 years, our Enthalpy Based Thermal Evolution of Loops (EBTEL) approach has been developed \citep{ebtel_2008,ebtel_2012a,cargill2015,barnes_2016a,barnes_2016b}.
The essence of EBTEL is that an impulsively-heated loop proceeds through three phases: first, in response to increasing coronal heating, an enhanced heat flux 
enters the TR, which responds by an upward mass flow into the corona, commonly referred to as "evaporation". Secondly, once enough plasma has evaporated, coronal radiative losses increase to a time when the radiative and conductive losses are roughly equal. Finally, as radiative losses become dominant, the corona drains through an enthalpy flux to the TR \citep{cargill2015}. EBTEL is a zero-dimensional model that solves for coronal averages of the temperature and density, with the TR responding to heat and enthalpy fluxes to and from the corona. In general it gives good agreement with a full 1D solution on a variety of problems \citep{ebtel_2012a,cargill2015}, one exception being the early evolution of very impulsive (10 sec) electron heating bursts \citep{barnes_2016a}. 

In this paper, we enhance the EBTEL model to include a variation in the cross-sectional area of a loop. While some studies \citep[e.g.][]{klim_area_2000} argue that there are also strong suggestions that the cross-sections of observationally distinct loops are roughly constant, it is clear that the magnetic field must diverge with height on average in the corona. The cross sections of observed loops may expand preferentially in the line-of-sight direction, in which case it would not be detected \citep{mal_schrijver}, though this idea has recently been questioned \citep{klim_def_2020}. Further, extrapolation of photospheric magnetograms sometimes gives large area changes as one goes from chromsphere to corona \citep[e.g.][]{mok_2008,asgari_2013}. In developing the EBTEL model to include this area change, it became apparent that the knowledge of the physics of non-uniform area in static loops was incomplete despite being discussed by a number of authors \citep[e.g.][]{vesecky_1979,levine_pye,rabin_1991,dudik_2009,martens_2010}. Thus a major part of this paper will address static loops, and in turn this defines the range of applicability for the EBTEL model.

In Section 2 we derive the EBTEL equations for a non-uniform area. Section 3 discusses static loop models, and Section 4 presents the new dynamic EBTEL results.  Appendix A addresses an additional approximation in EBTEL due to the non-uniform area and Appendix B discusses the useful analytic approach to static loops of \citet{levine_pye} and \citet{martens_2010}. 

\section{The EBTEL equations with an area variation}

The one-dimensional (along a field line with a coordinate {\it s}) energy equation with a variation in the cross-sectional area $A(s)$ is:
\begin{equation}
\frac{\partial E}{\partial t} = -\frac{1}{A(s)}\frac{\partial}{\partial s}\Bigg(A(s)v\left[ E + p \right] +A(s) F_c \Bigg) +Q -n^2\Lambda(T)
\end{equation}
in the usual notation with $E = p/(\gamma -1) + 1/2 \rho v^2$, $F_c = -\kappa_0 T^{5/2}dT/ds$ is the heat flux with $\kappa_0 = 8.12 \times 10^{-7}$ in c.g.s. units, $\Lambda(T)$ an optically thin radiative loss function \citep[e.g.][]{ebtel_2008} and $Q(s,t)$ an imposed heating function. There is also the equation of state for a fully-ionised electron-proton plasma: $p = 2nkT$. 

The EBTEL method assumes that the upper solar atmosphere can be split into two parts: a corona and a transition region (TR). The length of the combined corona and TR (usually referred to as the loop half-length) is defined as $L$, with $s = 0$ at the base of the TR and $s = L$ at the apex. The top of the TR (defined as $s = s_0$, with quantities there denoted by subscript ‘0’) is defined as the location where energy transport by thermal conduction changes from a loss to a gain. Assuming subsonic flows, we follow \citet{ebtel_2008} and \citet{ebtel_2012a} and integrate Eq (1) over the corona to obtain:
\begin{equation}
\frac{A_c L_c}{\gamma-1}\frac{dp_c}{dt} = A_0 \left[\frac{\gamma}{\gamma-1}v_0p_0 + F_{c0}\right] + A_c L_c \left[Q_c - R_c /L_c \right ]
\end{equation}
where subscript "c" denotes a coronal quantity so that $L_c$ is the distance from the top of the TR to the loop apex, and $A_c$ the average of the area in the coronal segment. $p_c$ is the average coronal pressure, with the pressure at the loop apex ($p_a$) calculated in a way that includes gravitational stratification \citep{ebtel_2012a} . The heating $Q$ is assumed to be spatially uniform. At the loop apex we impose symmetry conditions such that $v = F_c = 0$ there. The integral of the radiative losses can be written formally as $R_c = 1/A_c \int A(s) n^2 \Lambda(T) ds$ \citep{klim_luna_2019}, the spatial integral is from $s_0$ to the loop apex. Note that it has been assumed that the spatial integral of the left hand side of Eq (1) can be written as the product of the average coronal area and pressure. This is discussed further in Appendix A. 

Similarly, integrating over the TR gives:
\begin{equation}
\begin{split}
\frac{A_{TR} L_{TR}}{\gamma-1}\frac{dp_{TR}}{dt} = &-A_0\left[\frac{\gamma}{\gamma-1}v_0p_0 + F_{c0}\right]\\
&+A_{TR} L_{TR} \left[Q_{TR} -R_{TR}/L_{TR} \right ]
\end{split}
\end{equation}
with $v = F_c = 0$ imposed at the base of the TR, subscript $TR$ denotes a TR quantity and $R_{TR}$ is now an integral over the TR: $R_{TR} = 1/A_{TR} \int A(s) n^2 \Lambda(T) ds$. 

We now set $p_c = p_{TR} = p$ and add these two equations to get:
\begin{equation}
\begin{split}
\frac {\left [A_c L_c + A_{TR} L_{TR} \right ]}{\gamma -1}\frac{dp}{dt} =  &A_c L_c Q_c + A_{TR} L_{TR} Q_{TR} \\
&- \left[A_c R_c + A_{TR} R_{TR}\right ]
\end{split}
\end{equation}
Setting $Q_{TR} = Q_c = Q$ and defining $L^* = L_c + A_{TR}L_{TR}/A_c$, we obtain:
\begin{equation}
\frac {L^*}{\gamma -1}\frac{dp}{dt} =  L^* Q  - R_c (1 + C_1 A_{TR}/A_c)
\end{equation}
where $C_1 = R_{TR}/R_c$, as in our earlier work.
If we set $A_c = A_{TR}$ then we recover:
\begin{equation}
\frac {L}{\gamma -1}\frac{dp}{dt} =  LQ - R_c (1 + C_1)
\end{equation}
with $L = L_c + L_{TR}$, the EBTEL pressure equation from our earlier papers\footnote{Note that \citet{ebtel_2008} did not distinguish between $L$ and $L_c$.}. 

The equation for the coronal density is given by:
\begin{equation}
\frac{\partial n}{\partial t} =  -\frac{1}{A(s)}\frac{\partial}{\partial s} (nvA(s))
\end{equation}
which integrating over the corona and using the equation of state gives:
\begin{equation}
A_c L_c \frac{dn}{dt} =  n_0v_0A_0 = \frac{pv_0A_0}{2kT_0}
\end{equation}
where $n$ is now the coronal average. Again we have written the integral of the left hand of Eq (7) as the product of the average area and average density, as discussed in Appendix A.
We then use the TR energy equation (3) to solve for $pv_0A_0$ such that:
\begin{equation}
\frac{\gamma}{\gamma-1}A_0pv_0 = -\left [ A_0 F_{c0} +\frac{A_{TR}R_cL_c}{L^*} \left( C_1 - L_{TR}/L_c \right ) \right ]
\end{equation}
where Eq (5) is also used: the same result arises from using (2) and (5). This then gives:
\begin{equation}
A_c L_c \frac{dn}{dt} =  - \frac{\gamma - 1}{2k \gamma T_0} \left [ A_0 F_{c0} + A_{TR} R_c \frac{L_c}{L^*} \left( C_1  - \frac{L_{TR}}{L_c} \right ) \right ]
\end{equation}
Setting $A_c = A_0 = A_{TR}$ gives:
\begin{equation}
\frac{dn}{dt} =  - \frac{\gamma - 1}{2k \gamma T_0 L_c} \left [F_{c0} + R_c \frac{L_c}{L} \left( C_1  - \frac{L_{TR}}{L_c} \right ) \right ]
\end{equation}
which is the same as in the earlier papers except for the correction $L_{TR}/L_c$ on the right hand side due to the change in the TR pressure. For uniform area loops, $L_{TR}/L_c$ is of order 0.1 - 0.2, so is a small correction during most phases of evolution. However during radiative cooling $C_1$ may be $< 1$ \citep{ebtel_2012a}, so it could be significant and is retained in the modelling. Note also the presence of $L_c$ instead of $L$ in the leading rhs coefficient and also in the definition of $F_0 = -2/7\kappa_0 T_a^{7/2}/L_c$, where $T_a$ is the apex temperature. 

To solve these equations, we remove $T_0$ by defining two constants: $C_2 = T/T_a$ and $C_3 = T_0/T_a$, where T is now the coronal average, so that EBTEL solves:
\begin{equation}
\frac {1}{\gamma -1}\frac{dp}{dt} =   Q  - (L_c/L^*)n^2\Lambda (1 + C_1 A_{TR}/A_c)
\end{equation}
and
\begin{equation}
\frac{dn}{dt} =  - \frac{C_2(\gamma - 1)}{2k \gamma T L_c C_3} \left [ \frac{A_0}{A_c} F_{c0} + \frac{A_{TR}}{A_c} n^2 \Lambda \frac{L_c^2}{L^*} \left( C_1  - \frac{L_{TR}}{L_c} \right ) \right ]
\end{equation}
Two area ratios arise and the three constants $C_1$, $C_2$ and $C_3$ are discussed fully in our earlier work. For constant area, $C_2 = 0.9$ and $C_3 = 0.6$ at all times. $C_1$ is allowed to vary with time such that $C_1 = 2$ when the loop density reaches its maximum, $C_1 > 2$ when conduction dominates radiation \citep{barnes_2016a} and $C_1 < 2$ when radiation dominates \citep{ebtel_2012a}. $C_1$ is further modified when gravitational stratification is included \citep{ebtel_2012a}.

\section{Static Loop results}

Static loop models are of interest because (a) EBTEL relies on them for the determination of $C_1$ and (b) with the exception of \citet{vesecky_1979}, previous work has not addressed the changes in the physics due to area variations. Two approaches are used. One develops scaling laws based on the EBTEL equations and the second considers solutions to the static energy balance equation.

\subsection{Scaling laws}

We can write approximate scaling laws using Eq (12) and (13) in the static limit:
\begin{equation}
n^2 = (L^*/L_c)Q/[\Lambda(1 + C_1 A_{TR}/A_c)]
\end{equation}
\begin{equation}
T_a^{7/2} = (7L_c^2Q/2\kappa_0) (A_{TR}/A_0) \left [ \frac{ C_1  - L_{TR}/L_c}{1 + C_1 (A_{TR}/A_c)} \right ]
\end{equation}
or
\begin{equation}
Q = (2\kappa_0 T_a^{7/2}/7L_c^2)(A_{0}/A_{TR}) \left [ \frac{(1 + C_1 A_{TR}/A_c)}{C_1 - (L_{TR}/L_c)}\right ]
\end{equation}
\noindent
The recent scaling laws of \citet{klim_luna_2019} who also considered an area variation (Eq (5) and (11) of that paper), can be obtained by setting $L_{TR} = 0$ and $L_c = L$. It is assumed implicitly that $A_{TR} \le A_0 \le A_c$.

We return to the scaling laws in Section 3.5 but it is important to note that they are only approximate solutions to the energy equation. This should be contrasted with those presented by \citet{martens_2010} and discussed in Appendix B  which are an exact analytic solution, provided a single power law radiative loss function, and a cross-sectional area satisfying Eq (B1) are used. The difference lies primarily in the numerical coefficients rather than the relationship between quantities such as $T_a$, $Q$ and $L$. The exact solution represents the role of the entire atmospheric structure in balancing heating and radiation throughout the loop.

\subsection{Example with small and moderate area variations}

We consider the following normalised area profile (A(s)) in terms of the function f(s):
\begin{equation}
\begin{split}
A(s) = f(s), ~~~ &f(s) = 1 + f_1 sin^2 (\pi s/2s_a), ~~ s < s_a, ~~\\
&f(s) = 1 + f_1, ~~ s \ge s_a
\end{split}
\end{equation}
$s_a = L$ is a loop with a smooth area change along its entire length and by decreasing $s_a$ from $L$ to small values, we force the area change to be more localised at the loop base. Figure 1 shows $A(s)$ as $s_a$ increases from $0.05L$ to $0.95L$ for $f_1 = 4$. 
\begin{figure}
	\centering
		\includegraphics[width=\linewidth]{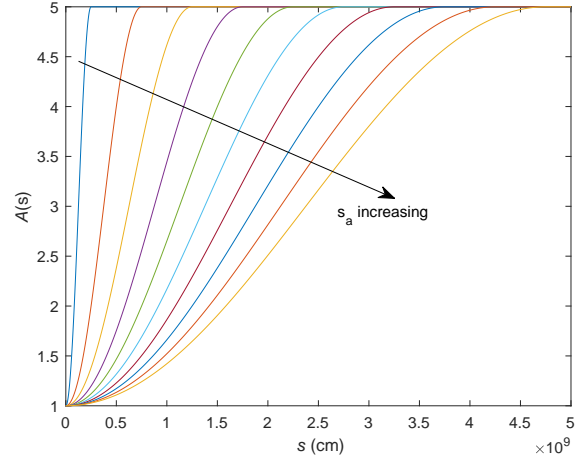}
	\label{fig:fig2_new}
	\caption{The normalised area profile given by Eq (17) as $s_a$ increases from 0.05L (leftmost curve) to 0.95L (rightmost).}
\end{figure}

We solve eq (1) numerically using a high-order Runge-Kutta scheme with $\partial / \partial t$ and $v$ set to zero and a prescribed cross-sectional area of the loop subject to a fixed chromospheric temperature at a point $s = 0$ (taken as $3 \times 10^4$K), and a vanishing heat flux at the loop apex ($s = L$). The heat flux at the base also vanishes. Solution of Eq (1) then requires specification of two of the following: $L$, $Q$, $T(s = L) = T_a$, $n(s = L) = n_a$ so that this is an eigenvalue problem \citep{martens_2010}, with the other two quantities determined by the need to satisfy the boundary conditions. The Sun will specify $Q$ and $L$, so that an iterative solution of (1) gives $T_a$ and $n_a$ once the boundary condition at the loop apex is satisfied. Alternatively, it is sometimes convenient to specify $T_a$ and $L$ and determine $Q$ and $n_a$. We specify $L$ and $Q$ and calculate $T_a$ and $n_a$\footnote{Despite its extensive use in the literature, we do not adopt the pressure as an output parameter. $T_a$ and $n_a$ are, in principle, measurable quantities, whereas $p$ is not.}.

The energy equation can be written in the following form: 
\begin{equation}
\frac{d}{ds}\left(\kappa_0 T^{5/2} \frac{dT}{ds} \right) +\kappa_0 T^{5/2}\frac{dT}{ds}\frac{1}{A(s)}\frac{dA}{ds}  +Q -n^2\Lambda(T) = 0
\end{equation}
For a loop with a monotonically increasing temperature and an area that increases from base to apex, the second term is always positive and so can be viewed as an effective "heating", as was noted by \citet{vesecky_1979}. What this means is that all else being equal, a loop with an area divergence will have a higher {\bf pressure} than one with uniform area: more "heating" does not necessarily imply a higher apex temperature. 

We begin by considering a loop of half-length 50 Mm, $s_a = L$, and vary the parameter $f_1$ between 0 and 5 so that the maximum apex area is six times larger than the base. $Q = 3.67 \times 10^{-4}$ ergs cm$^{-3}$ s$^{-1}$ which gives $T_a$ of order 2 MK. Gravity is neglected for the moment and a single power-law loss function of the form $\Lambda (T) = 1.95 \times 10^{-18} T^{-2/3}$ is used\footnote{The choice of a -2/3 power as opposed to the more usual -1/2 one is for consistency with our earlier work. Different coefficients lead to changes to the numerical values presented below, not to the underlying physics.}. Other loss functions are discussed later.  

Figures 2 and 3 show the results. The six panels of figure 2 show: $n_a$ in panel 1, $A_c$ (plus sign) and $A_0$ (circle) in panel 2, $T_a$ (star) and $T_0$ (circle) in panel 3, the ratio $C_1$ in panel 4, the ratio of conductive to radiative losses at $s = L$ in panel 5 and $L_{TR}/L$ in panel 6. For clarity $A_{TR}$ is not shown but increases from 1 to 1.2. $T_0$ is the temperature at the top of the TR and the averages $A_c$ and $A_{TR}$ are obtained a posteriori once the location of the top of the TR is determined. In panels 1, 3, 4 and 5, the black stars denote results when gravity is excluded. The upper two panels of Figure 3 show the conductive, radiative and heating terms in the energy equation as a function of distance for $f_1 = 0$ (left) and $1 + f_1 = 6$ (right). As the cross-sectional area at the apex increases we find (a) larger $n_a$, (b) slightly larger $T_a$, (c) a thicker transition region and (d) a slightly enhanced ratio of TR to coronal radiation per unit area, as represented by $C_1$. The first of these was also found by \citet{vesecky_1979}. 

\begin{figure}
	\centering
		\includegraphics[width=\linewidth]{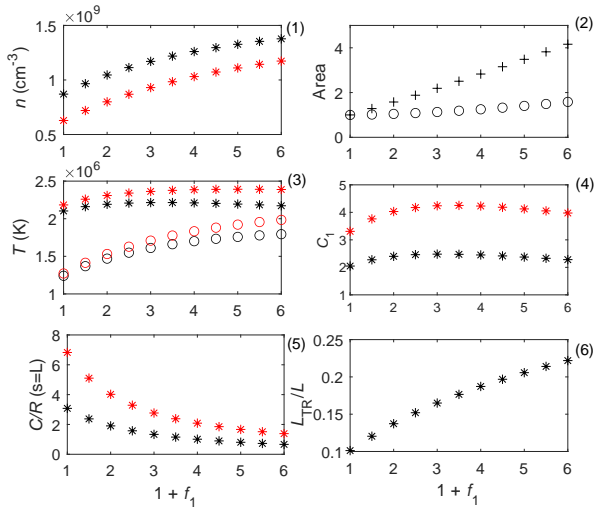}
	\label{fig:fig2_new}
	\caption{Loop properties showing the effect of increasing the cross-section at the loop apex: the horizontal axis is $1 + f_1$. The loop has 2L = 100 Mm and $Q = 3.67 \times 10^{-4}$ ergs cm$^{-3}$ s$^{-1}$. The six panels show (1) the apex density,(2) the area factors $A_c(+)$ and $A_0(o)$. $A_{TR}$ (not shown) increases from 1 to 1.2, (3) $T_a$ (stars) and $T_0$ (circles), (4) $C_1$, (5) the ratio of conductive (C) to radiative (R) losses at the apex ($C/R(s=L)$), and (6) the ratio of TR thickness to loop half-length. The black (red) symbols denote solutions where gravity is not (is) included.}
\end{figure}

\begin{figure}
	\centering
		\includegraphics[width=\linewidth]{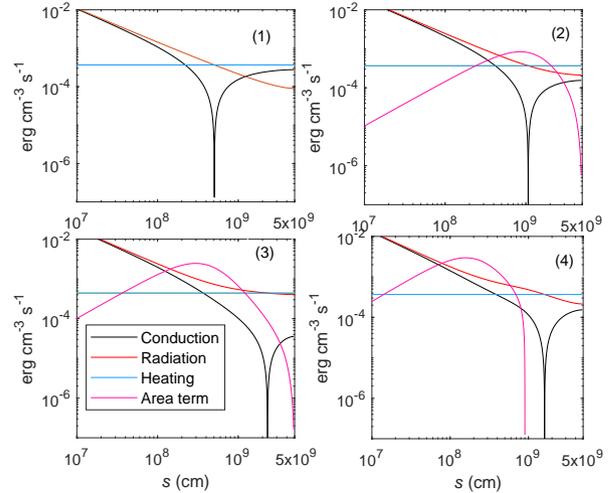}
	\label{fig:fig3_new}
	\caption{Magnitudes of terms in the energy equation for a loop with constant area (upper left: panel 1) and a loop with $f_1 = 5$ and $s_a = L$ (upper right: panel 2), in the absence of gravity. The lower plots show $f_1 = 50, s_a = L$ (left: panel 3) and $f_1 = 5, s_a = 0.2L$ (right: panel 4), and are discussed in Section 3.4. In each plot, conductive (black) and radiative (red) losses and heating (blue) are shown. Where shown, the pink curve represents an effective "heating" due to area divergence (see text).  Conduction is a loss (gain) to the right (left) of where it passes through zero. The loops all have 2L = 100 Mm and $Q = 3.67 \times 10^{-4}$ ergs cm$^{-3}$ s$^{-1}$.}
\end{figure}
\noindent

In the "standard" picture of loops, the role of the TR is to radiate away the total downward coronal heat flux (i.e. the sum of the heat flux over the loop cross-section). The TR is thin with $L_{TR}/L_c \sim 0.1 - 0.15$, and is the origin of 2/3 of the loop's radiation \citep{ebtel_2012a}. When gravitational stratification is included, the TR radiation predominates more. As the loop becomes constricted, the volume available in the TR to radiate away the total coronal heat flux diminishes. Thus to obtain equilibrium the TR and coronal density must both increase and/or the TR volume also increase. Figure 2 shows that both occur. This simple picture of an increase in coronal density and pressure and a thicker TR holds for all cases we consider. The loop temperature profile adjusts so that the higher density leads to the same radiative loss summed over the entire loop. Since the loss function decreases with temperature, more of the loop is at higher temperatures, and the profile $T(s)$ becomes flatter \citep[e.g.][]{vesecky_1979,martens_2010}. 

However, panel 5 of Figure 2 and Figure 3 show that the "standard" picture of loops begins to break down as $f_1$ increases. Considering Figure 3 first, the upper left panel shows a loop with constant cross-sectional area with conduction roughly equal to heating in the corona and conduction roughly equal to radiation in the TR. Increasing $f_1$ leads to a situation when conduction ceases to dominate the coronal energy balance, as seen in the upper right panel. The cause is the increase in the loop density such that coronal radiation becomes more important than conduction to the TR. This arises at roughly $f_1 = 3.5$. In terms of the temperature profile, the loop has become more "isothermal" in the corona, a well known effect of area constriction \citep{martens_2010}. Also, the "effective heating" (as defined earlier) is maximised at the top of the TR (top right panel of Figure 3) and dominates radiation there. 

Fixing $T_a$ and varying $Q$ leads to similar results, with differences in the exact numbers. Adopting the generalised radiative loss function of \citet{ebtel_2008} leads to the $C/R(s = L)$ ratio falling below unity at $f_1 = 4.5$. Finally, the red stars in Figure 2 show results when gravitational stratification is included. This keeps the apex $C/R(s = L)$ ratio above unity for all the $f_1$ considered here, but it falls from 6 to 2 as $f_1$ increases. Other quantities show expected variations (e.g. $C_1$ and $n_a$) or little change ($T_a$): the ratio $L_{TR}/L$ is not shown because the two plots overlap. 

We can also compare these results, especially the increase in loop pressure with $f_1$, with the TR studies of \citet{rabin_1991} who prescribe $A = A(T)$ (see Appendix B). This differs from the present work in that Rabin does not solve for the thermal structure of the entire atmosphere, but instead imposes a lower boundary condition on the heat flux, and iterate on the loop pressure until a fixed temperature is reached at a given height. The heat flux at that height is thus an output of the model. \citet{rabin_1991}, see his Figure 1, considers three TR area models that he calls "tee", "cone" and "bowl"\footnote{The "tee" geometry resembles the object used to elevate a ball in golf prior to hitting it into the water.}. The volume associated with each increases so that, based on our arguments above, for identical area profiles and heat fluxes at the top of the model, one would expect the tee to have a higher pressure than a cone which in turn has a higher pressure than a bowl. Further, larger area factors should have higher pressure than small ones. Comparison with the tee models is rendered difficult by the heat fluxes at the upper boundary differing by half an order of magnitude as the area factor changes. This implies that the coronal part of the loop differs between the cases. However, more constricted loops do show higher pressure. The bowl and cone cases do permit the desired comparison. As the area constriction increases in both, so does the pressure. And the cone does have higher pressures than the bowl. Thus this work agrees with our premise that what controls the loop conditions is the volume in the TR able to radiate away a downward heat flux.

\subsection{Large area variations}
Modelling of coronal magnetic fields based on photospheric magnetograms suggests that large area variations can arise. For example, \citet{mikic_2013} and \citet{froment_2018} considered a factor of 10 from chromosphere to apex, while \citet{asgari_2012,asgari_2013} consider factors in excess of 100 in longer loops. Here we consider much larger values of $f_1$. Figure 4 and the lower left panel of Figure 3 show the same quantities as Figure 2 and the upper panels of Figure 3, with the black (red) stars showing the results without (with) gravity. Without (with) gravity the ratio $C/R(s=L)$ falls below unity for $f_1 > 3 (7)$ and the TR increases to between 40 and 50 \% of the loop. Figure 3 shows how inconsequential conduction has become in the coronal part of the loop. This in turn implies that the coronal density should change little as $f_1$ becomes large because the coronal energy balance between heating and radiation is independent of the cross-sectional area. Figure 4 shows that this is indeed the case.

Thus the "standard" loop picture has entirely broken down for these large area variations. Further, in the no-gravity case, we see that $C_1$ approaches unity when $f_1$ exceeds 20. In these cases the loops are isothermal over most of their length and, with the upper boundary of the TR located at $0.4L$ or greater, the distinction between TR and corona becomes unclear, and the formal definition of the top of the TR given earlier is probably meaningless. A thin layer of steep temperature gradient still exists, but it is confined to near the footpoint, far below $L_{TR}$. It deviates substantially from our formal definition of transition region.

\begin{figure}
	\centering
		\includegraphics[width=\linewidth]{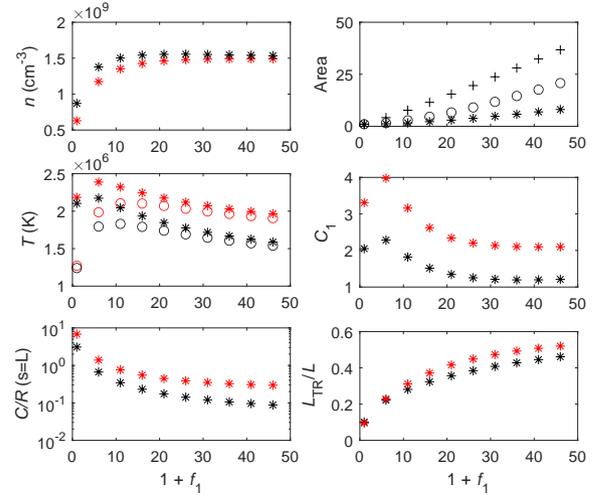}
	\label{fig:fig2_new}
	\caption{As Figure 2 except for larger values of $f_1$. In the upper right panel, the * symbol denotes $A_{TR}$. For clarity, the ratio $C/R$ is now shown on a log scale. As before, black (red) symbols denote solutions without (with) gravity included.}
\end{figure}

\subsection{Different area profiles}
We now consider what happens when $s_a$ is varied for the same range of $f_1$ with $s_a$ varying between 0.05L and 0.95L. Small values of $s_a$ localise the area variation to the lower part of the loop so that there is a larger volume in the TR available to radiate away the downward heat flux. Figure 5 shows the results for $f_1 = 5$ (stars) and $f_1 = 50$ (circles) in the same format as Figure 2. The lower right panel of Figure 3 shows the results for $s_a = 0.2L$ and $f_1 = 5$ so that the entire TR is constricted. There are relatively small changes in $n_a$, and $T_a$ and the area-associated "heating" now becomes strongly localised. In Figure 5 we see that the density increases with $s_a$, as expected when the loop becomes constricted over a greater part of its length. 
With the exception of the ratio $C_1$, there is relatively little change in the loop properties as $s_a$ increases. 

\begin{figure}
	\centering
		\includegraphics[width=\linewidth]{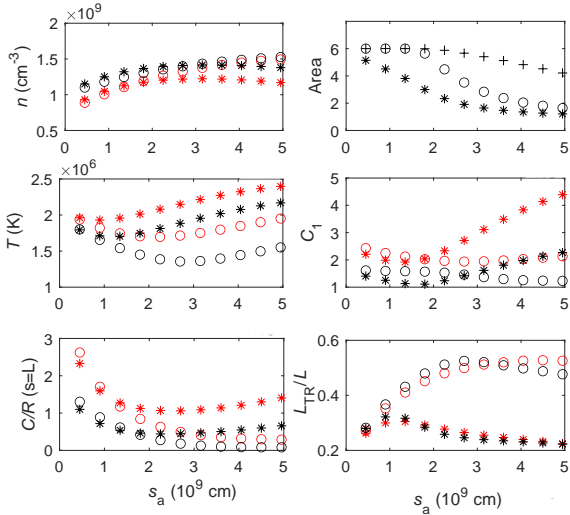}
	\label{fig:fig2_new}
	\caption{As Figure 2 except $s_a$ is allowed to vary. $f_1 = 5$ (stars) and $f_1 = 50$ (circles). Black (red) symbols ignore (include) gravity.}
\end{figure}
\subsection{Comparison with scaling laws}

We now compare the scaling laws described in Section 3.1 with exact solutions of the energy equation. Figure 6 shows the difference between the exact and scaling law solutions, normalised with respect to the exact solutions, so that for example $\Delta n = [n(exact)-n(scaling)]/n(exact)$. The top two rows show $\Delta n$ and $\Delta T$ respectively with $s_a =  L$ and $0 \le f_1 \le 5$ and $0 \le f_1 \le 50$ in the left and right columns. The third and fourth rows show $0.05L \le s_a \le L$ for $f_1 =  5$ and $f_1 = 50$ (left and right columns). 

The exact solutions and scaling laws can be compared in a number of ways. The black stars use scaling law values calculated with the numerical values of $C_1$, $L_c, L_{TR}$ and the various area factors obtained by the numerical solution of the energy equation. However, there are instances where a simpler approach is desirable when the scaling laws are implemented without knowledge of the detailed energy equation solution, as done in Eq (5) and (11) of \citet{klim_luna_2019}. To do this, we set $L_c = L, C_1 = 2$. These are the red stars in the panels of Figure 6. Finally, we use our scaling laws, with $L_c$ and $L_{TR}$ from the exact solutions, but with $C_1 = 2$, shown as the blue stars in Figure 6. 

We see that over all parameter ranges, the difference between the actual and scaling law densities is relatively small, at most of order 20\%. This arises because the density scaling law is a simple statement that the energy deposited must equal that radiated. The temperature that comes in via the loss function is a modest correction. Indeed the largest errors in the density arise for loops with uniform area. 

The temperature scaling law(s) perform less well. We see that the agreement for small $f_1$ is good given the assumptions used in obtaining the scaling laws. Indeed, if we increase the 2/7 factor to 3.25/7 in the approximation of the heat flux, the agreement between the temperatures becomes excellent.\footnote{It is instructive to compare the expression $(2A_0\kappa_0/7)T_a^{7/2}/L_c$ with the numerical value at the top of the TR. They differ by a factor two} However, as $f_1$ increases further, the discrepancies in the apex temperature become more marked. In particular the Klimchuk and Luna model shows significant deviation since the assumption of a thin transition region clearly breaks down. This could be attributed to the reduction of the heat flux to the TR. On the other hand, the premise of the scaling laws is violated for large $f_1$, namely that one cannot equate conduction and radiation in the loop. It seems as though the scaling laws should not be used once the coronal conductive losses fall much below the radiative ones. The earlier figures suggest that this is for $f_1 > 5$. This is also when $L_{TR}$ becomes a significant fraction of $L$.

\begin{figure}
	\centering
		\includegraphics[width=\linewidth]{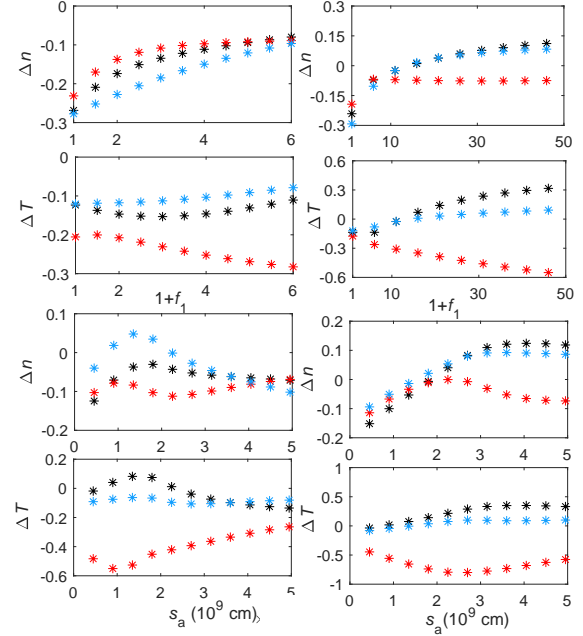}
	\label{fig:fig2_new}
	\caption{Difference between the full solution of the energy equation and scaling law values of $T$ and $n$, normalised wrt the full solution. Black, red and blue stars show, respectively, the full scaling laws, those of \citet{klim_luna_2019}, and the full laws, but with $C_1 = 2$. The upper panels show variations of $f_1$ with small (large) ranges of $f_1$. The lower panels show variations of $s_a$ for $f_1 = 5$ and 50 respectively. Gravity is excluded in all cases.}
\end{figure}

\section{Time-dependent solutions}

We now solve the time-dependent EBTEL equations (12) and (13) for a simple heating model. The results of the previous Section suggest that we can retain $C_1 = 2$, with the modifications of \citet{ebtel_2012a} provided the variation in area between base and apex is small enough, of order 5. This in turn implies that the ratio $L_{TR}/L$ is small and conduction is the dominant coronal loss mechanism up to the start of the radiative phase. These are essential assumptions of the time-dependent EBTEL model. Thus the parameters $C_1, C_2$ and $C_3$ are as in the earlier papers and the \citet{ebtel_2008} radiative losses are used. It is assumed that $L_{TR}/L = 0.15$. Figure 7 shows a case where the loop has 2L = 80, with a triangular pulse of duration 200 sec and peak 0.1 ergs cm$^{-3}$ s$^{-1}$. There is a background heating of $3 \times 10^{-5}$ ergs cm$^{-3}$ s$^{-1}$ to ensure that (a) the loop starts from an equilibrium and (b) during the cooling phase, negative temperatures and densities do not arise.  The four panels show the temperature, density, pressure and $T-n$ phase plane. The solid lines are the results for a constant area. We consider two extremes in area variation: one where $A_0 = A_{TR}$ (dashed), and one where $A_0$ is comparable to the coronal scale, $A_0 = 2A_{TR}$ (dotted). These are specified at $t = 0$, and remain unchanged as the loop evolves. We consider a factor 3 total divergence (i.e. $A_c = 3 A_{TR}$). The case with $A_0 = A_{TR}$ is probably the most realistic for a loop (i.e thin TR and mostly coronal field change), as discussed in \citet{guarrasi_2014}.
\begin{figure}
	\centering
		\includegraphics[width=\linewidth]{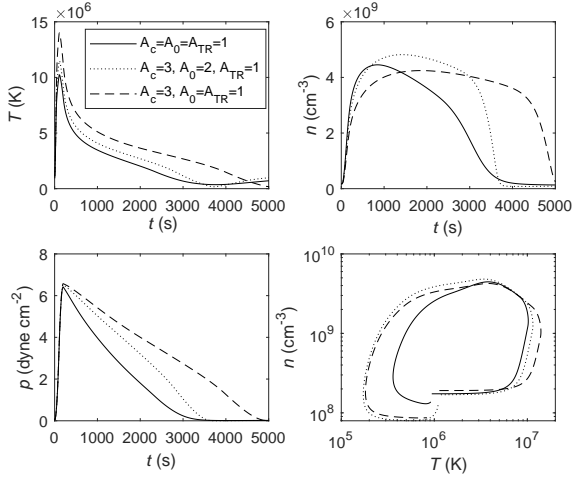}
	\label{ebtel_area}
	\caption{The evolution of a loop evaluated from the time-dependent EBTEL equations with uniform area (solid), $A_{TR}$, $A_0$ and $A_c$ = 1, 1, 3 (dashed) and 1, 2 and 3 (dotted).}
\end{figure}

 Even with such a relatively modest area change, the changes from uniform area are significant. In general terms, the rate of energy loss from the corona by either thermal conduction or enthalpy is less for a constricted loop than if there were no constriction. Consequently, the coronal temperature increases more quickly during the heating phase and decreases more slowly during the initial cooling phase. While the time of this peak temperature is similar for all cases, corresponding approximately to the peak of the heating, the highest temperatures arise when the area variation lies entirely above the TR. Here the heat flux into the TR is constricted by a factor three ($A_0/A_c$) compared to a factor of two when some area variation in the TR is allowed. 

The maximum density is similar in all cases, of order $4 \times 10^9 ~ $cm$^{-3}$, but the subsequent evolution differs considerably. The rise in density is determined by a competition between the downward heat flux and the (lack of) ability of the TR to radiate this away. Thus the longest delay in the maximum density with respect to temperature arises for the smallest TR volume, the dashed curve. The decline in density arises due to two effects: in situ radiation from the corona that reduces the gravitational scale height, and TR radiation powered by the downward enthalpy flux. The latter is limited by the TR volume, so this density decline is (initially) slower for the dashed curve. However, in both cases there is a significant difference from the uniform area case at large times. For uniform area, there is a smooth decline to small density. For the non-uniform area cases, there is a catastrophic decline, occurring after 3500 sec for $A_0 = 2 A_{TR}$ and after 4500 sec for $A_0 = A_{TR}$. This arises because in both cases the coronal density is held higher than is the case with constant area by the relative inefficiency of the downward enthalpy flux. This is a stronger effect for the narrow TR.  Eventually as the temperature falls, the high coronal density leads to overwhelming radiative losses and the loop cools catastrophically.

\subsection{Comparison between EBTEL and one-dimensional hydrodynamic simulations}

We now show a comparison of EBTEL simulations with results from the adaptive mesh one-dimensional hydrodynamic Hydrad code \citep{sb_pc2013,reep_2019}. The loop parameters and heating functions are as in Figure 7. In the Hydrad models the loop has a total length of 80 Mm (so that the half-length L = 40 Mm) to which is attached a stratified chromosphere at each footpoint with thickness 5 Mm. Note that L does not include the chromosphere.  Both Hydrad and EBTEL use the radiative losses of Klimchuk et al (2008).
Two (normalised) area models are considered:
\begin{eqnarray}
A(s) = (1 + tanh^q(\pi s/L_s))^{log3/log2}\\
A(s) = (1 + sin^q(\pi s/2L))^{log3/log2},
\end{eqnarray}
 which localise the area variation in the TR and corona respectively. In the tanh profile, $L_s$ = 20 Mm. The area at the loop apex is three times that at the top of the chromosphere. $A_{TR}$ and $A_0$ are calculated using these assumed profiles with $L_{TR} = 0.15L$ and are held constant in EBTEL as the loop evolves. For the tanh profile, the ratios $A_{TR}/A_c$ and $A_0/A_c$ decrease markedly as q increases while the sin profiles show little variation with q. The exact values in each simulation are stated in the figure captions but Eq (19) and (20) present differing challenges for a comparison between Hydrad and EBTEL. In terms of Figure 7, the sin profile is closest to the dashed lines and the tanh profile to the dotted lies, though the precise values of the areas are different.

Cases with q = 0 - 4 have been run for both area profiles. Figures 8 and 9 show results for the tanh and sin profiles respectively for q = 1 (upper panels) and 4 (lower panels). The cases with q = 2 and 3 give results that are intermediate between those shown.  The solid (dashed) lines show Hydrad (EBTEL) solutions. The red lines show values averaged over the coronal portion of the loop and the black lines those at the apex. For EBTEL, the coronal averages are as defined in Eq (12) and (13). For Hydrad the coronal averages are evaluated over the upper 85\% of the loop above the top of the initial model chromosphere (i.e. the top 34 Mm), as is the case with EBTEL under the assumption $L_{TR}/L = 0.15$. This ensures that we compare like-with like. Note also that the Hydrad averages are computed over the same spatial domain throughout the simulations.  For apex values, we average the Hydrad solutions over the top 20\% of the loop since past experience suggests that using precise apex values are overly noisy. For EBTEL, the apex temperature is related to the average by $T_a = T/C_2$, where $C_2= 0.89$ (Klimchuk et al, 2008, Cargill et al, 2012a). The apex density is related to the average using the formalism in Section 3.1 of Cargill et al (2012a) which accounts for gravitational stratification. [For reference, the Hydrad solutions for uniform area have a peak average temperature and density of 9.4 MK and $4.4 \times 10^9$ $\rm cm^{-3}$ respectively. The temperature falls below 1 MK at 2300 sec and the density to $2.5 \times 10^8$ $\rm cm^{-3}$ at 3600 sec. The density falls linearly as a function of time between its peak and this value.] 

\begin{figure}
	\centering
		\includegraphics[width=\linewidth]{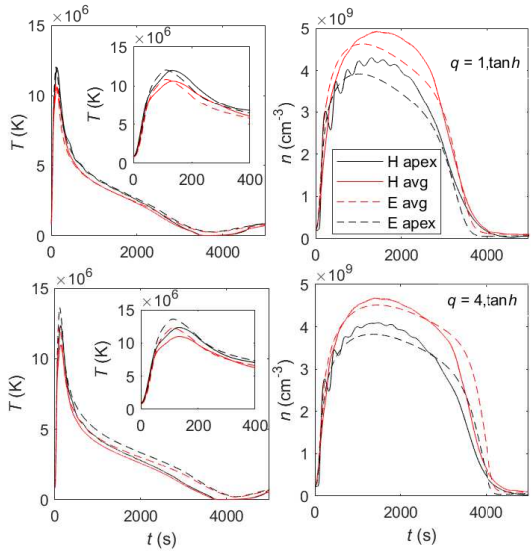}
	\label{fig:tanh_composite_nl}
	\caption{A comparison of Hydrad and EBTEL solutions of an impulsively heated loop with an area profile given by the tanh distributions. The red (black) lines are the average and apex values, with solid (dashed) being Hydrad (EBTEL), as summarised in the embedded box in the upper right panel. Temperature and density are shown as a function of time. The two upper (lower) panels show q = 1 and 4 respectively. In EBTEL, the upper panels have $A_{TR} = 1.72, A_0 = 2.38, A_c = 2.93$  and the lower ones $A_{TR} = 1.12, A_0 = 1.48, A_c = 2.80$. The sub-panel in the temperature plots shows the various temperatures for the first 400 seconds of the simulation.}
\end{figure}

\begin{figure}
	\centering
		\includegraphics[width=\linewidth]{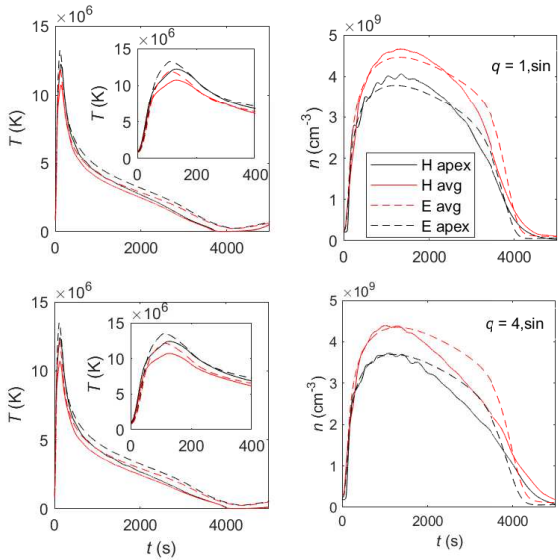}
	\label{fig:sin_composite_nl}
	\caption{As Figure 8 except the area profile given by the sin distributions. In EBTEL, $A_{TR} = 1.19, A_0 = 1.39, A_c = 2.4$ and the lower ones $A_{TR} = 1.0, A_0 = 1.0, A_c = 1.83$}
\end{figure}

In general, the Hydrad solutions for non-uniform area have the same generic properties as the EBTEL ones shown in the previous figure, namely a higher peak temperature, delay in the time of the maximum density, and an enhanced density throughout the radiative cooling phase when compared with the uniform area cases. Also, the comparison of the Hydrad and EBTEL temperatures (both apex and average) show a level of agreement comparable with our previous studies (see \citet{ebtel_2012a,cargill2015,barnes_2016a}). As the area constriction increases, the peak temperatures obtained by both EBTEL and Hydrad increase, as expected from Figure 7. The peak EBTEL values exceed the peak Hydrad ones by 1 – 2 MK, a percentage difference that is of the same order as we found for uniform area \citep{ebtel_2012a}. This higher peak then leads to slightly higher values of the EBTEL temperatures throughout the decay phase, though the rate of temperature decline is very similar in EBTEL and Hydrad.

As noted in \citet{ebtel_2012b}, obtaining good agreement between the density evolution in approximate and exact numerical models is more challenging. For uniform area, there is a tendency for the EBTEL density to exceed the Hydrad one, the value differing from case-to-case. In all cases shown here, the Hydrad peak density now exceeds the EBTEL one, though in all but the tanh area profile with q = 1, by a small amount. Even in this case, the excess is of order a few \%. In the draining phase, in all cases Hydrad and EBTEL show sustained higher densities than for uniform area. This general agreement of the densities between the two methods indicates that EBTEL is getting the important process of the TR response to a strong downward heat flux correct for these non-uniform areas. It is the difficulty of modelling this process with one-dimensional hydrodynamic codes without adequate numerical resolution that has been a primary motivation for our development of EBTEL.

Looking at the tanh cases, we expect $q = 4$ to have a more sustained high density than $q = 1$ since the TR is more constricted and this is what we find. For the sin cases, this effect is still present but less noticeable since the change in constriction as q increases is smaller. We also note that in the final stages Hydrad does not see as dramatic a catastrophic draining as EBTEL. One possible cause is that Hydrad may be better equipped to sustain a low-temperature hydrostatic equilibrium than EBTEL due to a pile-up of plasma at the footpoints, or a different form of cooling such as that discussed in \citet{pc_sb2013}, and not modelled by EBTEL, is operating.  

An important observational diagnostic of the heating in the core of active region loops is the temperature dependence of the emission measure ($EM(T)$) which scales in the range $T^2$ to $T^4$ for temperatures below 3 MK \citep{warren_2012,cargill_2014,barnes_2019}. For a single impulsive heating event, or a nanoflare train with well-separated heating bursts, at such temperatures the core of the active region loop is in the radiative cooling phase. In this regime \citet{cargill_1994} and \citet{pc_jk2004} showed that $EM(T) \simeq n^2 \tau_{rad}$ where $\tau_{rad}$ is the radiative cooling time at a given instant, defined as $\tau_{rad} \simeq 3kT^{1-\alpha}/\chi n$ for a power law radiative loss function of the form $\Lambda(T) = \chi T^\alpha$ . Thus $EM(T) \simeq n T^{1-\alpha}$ which for the commonly-used value $\alpha = -1/2$ gives $EM(T) \simeq n T^{3/2}$. When the loop area is uniform, the radiative/enthalpy cooling phase has $T \simeq n^2$ \citep{cargill_1995} so that $EM(T) \simeq T^2$ \citep[e.g.][] {cargill_2014}. With an area variation, Figures 7 - 9 suggest that the density remains higher in this cooling phase than for constant area.  Taking the extreme case of constant-density cooling, setting $n$ as a constant in the above expressions gives $EM(T) \simeq T^{3/2}$. Thus, despite the different behaviour of the density in the radiative cooling phase, the temperature dependence of $EM(T)$ shows little change in the presence of the modest area variations we consider, and is almost certainly not observable \citep{guennou_2013}. Note also that for loops with non-uniform area, the enhanced density in the radiative phase implies a higher value of the coronal  emission measure $EM(T)$ at a given temperature.

\section{Conclusions}

We have presented models that discuss the role of a non-uniform cross-sectional area in static and dynamic coronal loops. The results in all stages can be understood in simple terms that considers the response of the radiative losses from a constricted transition region to heat and enthalpy fluxes from the corona. For static loops, the smaller TR area leads to a higher coronal densities and a broader TR so that the downward heat flux may be radiated away. For impulsively-heated loops, the constricted TR leads to higher coronal temperatures during the heating phase and a sustained high density followed by rapid cooling in the radiative phase when the TR is unable to radiate the downward enthalpy flux. 

For large area variations, the standard picture of a static loop breaks down, with the coronal energy balance being primarily between heating and radiation as opposed to between primarily heating and downward conduction. These results suggest that caution is needed in modelling such loops with conventional concepts of loop energetics, as shown by the failure of the temperature scaling laws.

Since the EBTEL model makes use of some results from static loop models, we are able to set constraints on when area variations can be included in EBTEL. If we require $L_{TR}$ to be small compared with $L$ and conductive losses dominate during the heating and initial cooling phase, then EBTEL is limited to quite modest area variations, typically a factor four between base and apex. Nonetheless, the EBTEL results, and the comparison with Hydrad, indicate clearly the different physics to be expected in loops with non-uniform areas, although full (and computationally expensive) one-dimensional simulations will be required to verify this for large area variations.

\appendix
\section{A note on the average density and pressure in a loop with non-uniform cross-section}
While the average coronal density is clearly defined for a loop with uniform cross-section, a more detailed investigation is required for non-uniform area. For the mass equation (7), integrating the left hand side over the coronal loop portion formally gives:
\begin{equation}
L_c A_c \frac{dn_w}{dt} = n_0A_0v_0
\end{equation}
where
\begin{equation}
    n_w = \frac{1}{A_c L_c} \int A(s) n(s) ds.
\end{equation}
is the area-weighted average density. For constant area, $n$ and $n_w$ are the same, but for non-uniform area $n_w$ requires spatial information about the density and area profiles. In EBTEL we assume that $n_w = n$, where $n$ is the average density, $(1/L_c)\int n ds$. 

Justification for this can be addressed by comparing $n_w$ and n in hydrostatic loops: hydrostatic loop models underpin much of EBTEL due to the assumption of subsonic flows \citep{ebtel_2008,ebtel_2012a}. We have calculated $n$ and $n_w$ for all the static loop models considered in Section 3. We find that the ratio $n/n_w$ satisfies $1 < n/n_w < 1.05$ in all cases. The maximum of this ratio does not arise for the maximum $f_1$. As $f_1$ increases two effects arise. The coronal portion of the loop becomes more isothermal (see Figure 3), so that the density profile is also flatter. Secondly, the actual extent of the corona decreases as the ratio $L_c/L$ decreases, as seen in Figures 2 - 5. In both cases, the difference between $n_w$ and $n$ will then decrease. Thus the small discrepancy between $n_w$ and $n$ for large $f_1$ is somewhat artificial since the basic assumptions needed for EBTEL (in particular a narrow TR) are violated, as discussed elsewhere. 

We also examined a much longer loop with 2L = 400 Mm and a peak temperature of order 1.5 MK sustained by a heating of $10^{-5}$ ergs cm$^{-3}$ s$^{-1}$. Such lengths are the longest used in contemporary models (e.g. Asghari-Tahari et al, 2013; Froment et al, 2018). In this case the ratio $n/n_w$ rises to 1.14 for some values of $f_1$. As with the shorter loops, the maximum value of the ratio does not occur for the largest $f_1$, but for an intermediate value, $f_1 = 5$. So caution is warranted using EBTEL during evolution of long loops with non-uniform cross-sections at low temperatures.

We can also define an area-weighted pressure in the same way. The ratio $p/p_w$, where p is now the average pressure is closer to unity than $n/n_w$, of order 1.03 for the shorter loops and 1.1 for the longer. 

Two further points should be made. One is that EBTEL assumes a smooth variation of the plasma parameters throughout the loop, with the temperature and density being not far removed from that expected in a hydrostatic state: this is equivalent to our assumption of subsonic flows. Thus EBTEL cannot model cases involving localised plasma clumping or cooling when $n_w$ and n may differ considerably. Secondly, as the temperature increases during impulsive heating and subsequent cooling, gravitational stratification becomes less important and the ratio $n/n_w$ decreases towards unity.


\section{Analytic solutions for static loops}
It was pointed out by \citet{levine_pye} and \citet{martens_2010} that the assumption 
\begin{equation}
A(s)/A_a = (T(s)/T_a)^\delta
\end{equation}
permitted analytic solutions of the energy equation for a static loop without gravity and with a radiative loss function that is a single power law over all temperatures: $\Lambda(T)= \chi T^\alpha$, where subscript “a” corresponds to a quantity at the loop apex. We discuss the limitations of the assumption in Eq (B1) later, but the analytic solutions provide valuable guidance for more general and realistic area profiles discussed in Section 3. We follow the analysis of \citet{martens_2010}, see also \citet{kuin_martens}, and define the variable $\eta = (T/T_a)^{7/2+\delta}$. For a static loop and spatially constant heating, retaining the notation of Martens, the static energy equation is then:
\begin{equation}
\begin{split}
 &\epsilon \frac{d^2\eta}{ds^2}=\eta^\mu-\xi\eta^\nu, ~~ \mu = -\frac{(2-\alpha-\delta)}
{7+2\delta}, ~~\nu = \frac{2\delta}{7+2\delta} ,\\ 
&\xi=QT_a^{2-\alpha}/(p^2 \chi),  
\epsilon = \frac{\kappa_0T_a^{11/2-\alpha}}{(7/2+\delta)p^2L^2\chi}
\end{split}
\end{equation}
\noindent
where the pressure is constant. Defining the parameter $\lambda = (3/2+\alpha)/(2(2-\alpha))$, the energy equation is solved for a variable $u = \eta^{\nu-\mu}$ as:
\begin{equation}
s/L = \beta_r(u, \lambda+1, 1/2)
\end{equation}
\noindent
where $\beta_r$ is the normalised incomplete beta function and $\mu -\nu = -(2 - \alpha)/(7/2 +\delta)$. Applying appropriate boundary conditions at loop base and apex eliminates $\xi$ and $\epsilon$ from Eq (A2) and gives the scaling laws:
\begin{equation}
\begin{split}
&Q=\frac{p^2 \chi (7/2+2\delta)}{T_a^{2-\alpha}(3/2+2\delta+\alpha)}, ~~~\\
&pL = T_a^{(11-2\alpha)/4}\left(\frac{\kappa_0}{\chi}\right)^{1/2}
\frac{(3+4\delta+2\alpha)^{1/2}B(\lambda+1,1/2)}{4-2\alpha}
\end{split}
\end{equation} 
where B(a,b) is the beta function\footnote{Eq (48) – (50) of Martens have a number of typos, corrected here for uniform heating. \citet{bray_1991} also provide the correct scalings.}. Removing the pressure, these can be rewritten to give an expression for $n_a$ and $T_a$, analagous to Eq (14) and 15): 
\begin{equation}
\begin{split}
&T_a^{7/2} = (7L^2/2\kappa_0)Q \frac{2(2-\alpha)^2}{(1+4\delta/7)B(\lambda+1,1/2)^2} ,\\
&n^2 = \frac{Q(3/2+2\delta+\alpha)}{\Lambda(T)(7/2+\delta)}
\end{split}
\end{equation}
\noindent
Inverting u to obtain T, we find $T \sim 1/u^{(7/2+\delta)(\nu-\mu)} \sim u^{(2-\alpha)}$, so that, given the prescription (B1), the spatial structure of the temperature for a given value of $T_a$ is independent of $\delta$. This is confirmed by numerical solutions of the energy equation\footnote{Note that while Figure 4 of \citet{martens_2010} shows T(s) differing between uniform and variable loop cross-section, a heating function scaling as $T^{-3/2}$, as given by classical Ohmic heating, is used. }. Following \citet{ebtel_2012a} we can also evaluate the temperature at the top of the TR ($T_0$). This occurs when the right hand side of (B2) vanishes:
\begin{equation}
\frac{T_0}{T_a} = \left[\frac{7/2+\delta}{3/2+2\delta+\alpha}\right]^{-1/(2-\alpha)}
\end{equation}
so that as $\delta$ increases, $T_0/T_a$ also increases. [The analytic solution for $C_1$, defined earlier, and discussed by \citet{ebtel_2012a} cannot be repeated when $\delta > 0$.]

We have solved Eq (B2) for a range of area profiles, defining $\delta$ as $\delta = log_{10}(A_{max})/log_{10}(T_a/10^5)$. $A_{max} = A_a/A(T = 10^5)$ ranges from 1 to 10 so that $\delta$ varies between 0 and 0.77. A comparison of the solution of Eq (B2) with that of Eq (1) shows excellent agreement\footnote{In order to obtain agreement with $C_1$, a base temperature of almost zero (100 K) is required in the numerical solution.}. For a loop with $\alpha = -1/2$, $L = 50 Mm$ and $T_a = 2 MK$, we find that as $\delta$ increases, $C_1$ increases from 1.72 to 2.69, $T_0/T_a$ from 0.61 to 0.76 and $L_{TR}/L$ from 0.11 to 0.24. These analytic solutions reproduce the trends shown in Section 3.

In closing, we note that despite permitting analytic solutions, the area-temperature relationship is highly artificial, even in the more general formalism introduced by \citet{rabin_1991}. A credible scenario for coronal plasma structure is that the large-scale magnetic field of, for example, an active region is determined by the (global) dynamo process, within which smaller-scale processes provide the heating. Within this active region are many flux elements with cross-sectional areas A(s) given by the large-scale magnetic field. For a static loop, T(s) is determined by the solution of the energy equation for this prescribed A(s), and T(s) will not satisfy (B1) for any but the most serendipitous situations. It might be argued that the area in (B1) adjusts to the calculated temperature profile, but in a low-beta coronal plasma such a scenario is not credible, as was demonstrated in the simulations of \citet{guarrasi_2014}. These problems become more severe in dynamic loops discussed in Section 4.

\section*{acknowledgements}
We thank the referee for many helpful comments.
The work of JAK was supported the Internal Scientist Funding Model at GSFC (competitive work package program). WTB was supported by NASA’s Hinode program through the National Research Council. Hinode is a Japanese mission developed and launched by ISAS/JAXA with NAOJ as a domestic partner and NASA and STFC (UK) as international partners. It is operated by these agencies in cooperation with ESA and NSC (Norway).

\section*{Data Availability}
The EBTEL IDL code, including the modifications for non-constant cross-sections, is freely available and can be downloaded at https://github.com/rice-solar-physics/EBTEL.

\bibliographystyle{mnras} 
\bibliography{area.bib}

\bsp	
\label{lastpage}
\end{document}